# Developing a Unified Verification and Validation Activity Standard at JPL

Maximilian Vierlboeck, Ph.D.
Jet Propulsion Laboratory,
California Institute of Technology
4800 Oak Grove Dr.
Pasadena, CA 91109
maximilian.vierlboeck@jpl.nasa.gov

Ellen Van Wyk
Jet Propulsion Laboratory,
California Institute of Technology
4800 Oak Grove Dr.
Pasadena, CA 91109
ellen.j.van.wyk@jpl.nasa.gov

Natalia Sanchez
Jet Propulsion Laboratory,
California Institute of Technology
4800 Oak Grove Dr.
Pasadena, CA 91109
natalia.sanchez@jpl.nasa.gov

Bogdan Oaida
Jet Propulsion Laboratory,
California Institute of Technology
4800 Oak Grove Dr.
Pasadena, CA 91109
bogdan@jpl.nasa.gov

Torrance James Eberhart
Jet Propulsion Laboratory,
California Institute of Technology
4800 Oak Grove Dr.
Pasadena, CA 91109
torrance.j.eberhart@jpl.nasa.gov

Marie Piette Gomez
Jet Propulsion Laboratory,
California Institute of Technology
4800 Oak Grove Dr.
Pasadena, CA 91109
marie.y.piette@jpl.nasa.gov

Christopher Delp
Jet Propulsion Laboratory,
California Institute of Technology
4800 Oak Grove Dr.
Pasadena, CA 91109
christopher.l.delp@jpl.nasa.gov

Marijke G Jorritsma
Jet Propulsion Laboratory,
California Institute of Technology
4800 Oak Grove Dr.
Pasadena, CA 91109
marijke.g.jorritsma@jpl.nasa.gov

Alex Desharnais
Jet Propulsion Laboratory,
California Institute of Technology
4800 Oak Grove Dr.
Pasadena, CA 91109
alexander.h.desharnais@jpl.nasa.gov

**Abstract — Verification and validation practices (V&V) at NASA's Jet Propulsion Laboratory (JPL) have diverged over the past decade, creating fragmentation that increases overhead, reduces cross-project efficiencies, and inhibits institutional knowledge transfer. We present a unified V&V activity schema developed through human-centered design workshops involving 29 practitioners across multiple mission types and disciplines. The schema builds on a relationship-based architecture that allows for separating methods (Test, Analysis, Inspection, Demonstration, and Review of Design) while maintaining a common attribute set. Formalized as a platform-agnostic SysML model, the schema defines bidirectional relationships between requirements, V&V activities, venues, and evidence. Implementation in JPL's Jama platform demonstrates controlled customization through templates and modular item types, balancing rigor with agility while enabling automations, pattern reuse, and digital thread integration.**

**TABLE OF CONTENTS**

## 1. INTRODUCTION

The Jet Propulsion Laboratory (JPL) has designed, launched, and operated systems that venture into environments and conditions never before experienced by human-made artifacts. The development of such systems, while sharing methodologies with commercial system development, faces stark differences in operational constraints. Consider, for instance, that an automotive vehicle can be recalled, repaired, or patched after delivery—while far from ideal, this possibility enables fundamentally different approaches to risk management compared to systems that become physically inaccessible after launch.

This operational reality is further complicated by the novelty of each mission, which leads to methodologies that are often tailored to specific project needs and team preferences. While such customization has merit for addressing unique mission requirements, it has resulted in fragmented approaches across disciplines and projects, limiting reusability, scalability, and cross-project learning. The rigorous approach of traditional Verification and Validation (V&V) methodologies often requires significant resources and schedule considerations, thus making it all the more important that it be executed in an efficient and thoughtful manner.

Despite these unique constraints, today's aerospace landscape demands faster development times, more innovative solutions, multi-purpose systems, and higher cost-efficiency—driven not only by increased competition from commercially-oriented companies, but also by evolving mission requirements and budget realities. These pressures create a fundamental tension: how do we maintain the rigor necessary for mission-critical systems while achieving the development agility required in today's competitive environment?

This paper presents the development of a unified V&V Activity (VA) standard at JPL, designed to address the increasing divergence in implementation approaches. It might be reasonable to assume that a standard approach to implementing the V&V program already existed at JPL, and on the surface, that is certainly the case, as will be discussed in the subsequent section. However, there has been broad agreement within V&V practitioners at JPL that over the past decade or so there has been a significant increase in the variety of implementations deployed at the 'rubber-meets-the-road' layer.

Through a human-centered design approach involving practitioners and function owners across multiple disciplines and projects of various sizes, we developed a schema that maintains the necessary rigor for mission-critical systems while providing a 'common denominator' suitable for both high-complexity flagship missions, and low-cost, high-risk ones. The resulting framework enables more efficient developments through standardization without sacrificing the flexibility needed for mission-specific environments.

The primary contributions of this work include: (1) a unified V&V Activity schema that bridges previously fragmented approaches, (2) a relationship-based architecture that enables cross-discipline traceability and reuse based on previous successes, (3) implementation strategies for integration with a requirements management infrastructure, and (4) initial deployments on flight projects.

The remainder of this paper is structured as follows: Following this introduction, Chapter 2 provides an overview of the current state of practice for V&V activities at JPL, identifying the specific challenges arising from discipline-specific approaches and the opportunities for unification. Chapter 3 introduces the fundamental building blocks of the schema, defining the item types and relationship structures that form the theoretical foundation for the V&V solution. Chapter 4 details the human-centered design process employed, demonstrating how practitioner use cases, and operational needs drove the development rather than relying solely on prescribed best practices. Chapter 5 presents the resulting data model and information architecture, establishing both the practical implementation framework and the semantic foundation for future evolution of the standard. Chapter 6 addresses the specific implementation within Jama, JPL's institutional requirements management platform, including technical considerations and integration strategies. Chapter 7 describes pilot implementations on active flight projects, presenting initial results and lessons learned from early adopters. Finally, Chapter 8 provides a summary of contributions, their implications for V&V practice at JPL and the broader aerospace community, and future work opportunities before Chapter 9 concludes the paper.

## 2. State of Practice Overview

V&V represents a fundamental pillar of systems engineering, providing the framework through which engineered systems are demonstrated to meet their specified requirements and intended purpose. As such, V&V is closely tied to System Integration and also poses one of the main qualification steps for operation. In the V-Model of Systems Engineering, V&V forms the right ascending branch, systematically confirming that each level of system decomposition has been correctly implemented and integrated, ultimately demonstrating that the complete system satisfies stakeholder needs and operational requirements [1, 2].

While often discussed as a unified concept, V&V encompasses two distinct but complementary processes that together ensure system correctness and suitability. These two aspects as well as the respective details are outlined here before the setup and situation at JPL specifically is discussed.

### 2.1 Fundamental Definitions & Distinctions

While often discussed as a unified concept, V&V encompasses two distinct but complementary processes that together ensure system correctness and suitability. The distinction between verification ('Did we build the system right?') and validation ('Did we build the right system?') is well-established in systems engineering literature [1-6]. Both employ various methods including analysis, demonstration, inspection, and test, with selection based on trade-offs between confidence, cost, schedule, and technical risk [7]. V&V activities form a bidirectional flow between requirements and evidence, creating what Buede and Miller refer to as a 'chain' [8], with high-level requirements decomposed into lower-level specifications (related by Forsberg et al. [2] to Royce's Waterfall [9]). Traceability serves as the essential mechanism enabling both information flow and impact assessment across organizational boundaries [10, 11]. As ISO/IEC/IEEE 15288 acknowledges, V&V 'shall be implemented in accordance with applicable organization policies and procedures' [3], recognizing that standards may need adaptation for specific organizational contexts.

### 2.2 Verification and Validation at JPL

At a high level, the Verification & Validation approach employed at JPL follows the standard V-Model paradigm common in aerospace systems [1, 2]. JPL command media codifies the expectations of implementing the V&V program in the Institutional Project Verification & Validation Plan [12], which provides a template for projects to customize according to their specific circumstances. Foundational to this approach is the fact that each requirement must be formally and explicitly verified. The mechanics of exactly how that is achieved in practice, however, is largely beyond the scope of that document. So long as an independent auditor could, in theory, find the evidence that led to a given requirement's closure, the information architecture used by a project to achieve that end goal was left unspecified.

Despite starting from a seemingly common point, the evidence supplied by the past decade's worth of flight projects tells a surprising story: nearly every project took a different approach to how they managed the day-to-day aspects of V&V-related tasks. This was nowhere more evident than in the type and level of information stored in DOORS NextGen (DNG), JPL's requirements management database from mid 2010s until 2023: 24 instrument and full mission projects employed a total of 511 unique fields (e.g. ID, text, level, owner, etc.) across 20 unique V&V artifact types (e.g. Verification Activity, Verification Item, Test Case, etc.) to document all aspects of the V&V lifecycle. Furthermore, these numbers do *not* include the accounting of the pieces of information stored with the requirements themselves for the purposes of actual requirement closeout. At the very least, this suggests that the information management architecture within DNG was highly fragmented. That is likely a symptom of a larger underlying issue: lack of uniformity of practice. Simply put, the V&V state of practice at JPL had become highly disjointed.

It is natural to wonder how the state of practice could have devolved to such a fragmented one. A comprehensive answer to that question is beyond the scope of this paper, but there are two aspects that likely played a significant role that are relevant to our topic.

The first was the transition from DOORS Classic to DNG in the mid 2010s. DOORS Classic did not provide the opportunity for much project-to-project customization. Additionally, at the time, JPL had a well-established curriculum that trained users not just on the mechanics of how to use DOORS but how to implement the requirements lifecycle process in the tool. Both aspects amounted to, more or less, uniformity of practice amount individuals and projects. When DNG replaced its Classic sibling, the training curriculum was not updated. That, coupled with the customization flexibility DNG offered, enabled projects to essentially 'choose their own adventure'. While from the project and user perspective, the ability to tailor the implementation to the specific needs of a particular project was a big improvement, from an enterprise perspective, it led to the proliferation of artifact and field types to such a degree that managing DNG became both expensive and unwieldy. The proverbial wheel had been reinvented too many times.

The second reason is a more philosophical one in nature. Over time two schools of thought had evolved within the V&V practitioners at JPL, one that utilized the concept of V&V Activities as the main vehicle for planning, executing, tracking, and documenting all aspects of the V&V lifecycle, and a second one that relied more heavily on the requirements objects themselves to track the V&V and closure evidence. Essentially, there was a V&V activity-centric approach and a requirement-centric one. Necessarily, the information architecture needed to support the two approaches was different. However, the real problem was that even within each paradigm, there was little-to-no uniformity of practice, which each project largely defining its own DNG implementation, highly tailored to its needs.

Fundamentally, the challenge over the past decade has not been one of no longer knowing how to do V&V at JPL, but rather one of lacking discipline and consistency from project-to-project. In the aggregate, JPL's V&V information architecture had experienced what ecologist Daniel Pauly referred to as the Shifting Baseline Syndrome [13]. Adapted to our context, it is the tendency of each new project to accept the state of practice ushered in by its practitioners as the 'normal' or baseline condition, leading to 'generational amnesia': each group's perception of a 'healthy' state of practice is based on their own experiences, which may be a significantly degraded version of what a previous group witnessed. As a result, projects tended to be unaware of the extent of long-term degradation, which ultimately manifested as increased inefficiency and overhead in operating and maintaining the V&V information infrastructure.

The decision to migrate from DNG to Jama as the institutionally supported requirements management environment afforded the opportunity to stem the divergence of practice and re-vector towards a unified approach [13]. In 2024, a group of V&V practitioners initiated a grass-roots effort to identify potential changes to the state of practice. That effort, over the course of a year, evolved into a formal, structured one, eventually resulting into the standard described in the remainder of the paper.

## 3. DEVELOPMENT & HUMAN-CENTERED DESIGN

Building on the grassroots effort initiated in 2024, the team employed a human-centered design approach to develop a unified V&V Activity standard. This approach was chosen specifically to ensure the resulting schema would address the practical needs of diverse stakeholders while maintaining the rigor required for mission-critical systems. By bringing together practitioners from across disciplines and project types, the process aimed to create a solution that would both standardize practices and accommodate the legitimate flexibility needs identified in Section 2. The following sections detail this development process and the resulting schema architecture.

### 3.1 Background and Methodology

The study team made the choice to focus on a unified standard around the VA-centric paradigm (as opposed to the requirement-centric one). This was done for several reasons: 1) the VA-centric approach has deep roots at JPL and traces its origins to the DOORS Classic implementation; 2) the majority of the practitioners were already familiar with the approach, at least in theory 3) it was believed to enable an architecture that did *not* preclude a project from pursuing the requirements-centric approach if it so chooses.

Given the broad range of projects, perspectives, and roles that a standard VA schema needs to support at JPL, the design team chose to pursue a series of participatory workshops with JPL's V&V community using the Double Diamond model [14, 15].

The VA schema (V&V Activity schema) refers to the standardized information structure that defines how activities are documented, managed, and connected within JPL's requirements management platform. The schema encompasses the item types, attributes, relationships, and rules that govern VAs throughout their lifecycle. This schema serves as the foundation for transforming isolated documentation into an integrated information architecture that supports planning, execution, and closure of VAs across disciplines. The Minimum Viable Package (MVP) represents the core implementation of this schema containing the essential elements required for functionality while allowing for future expansion based on user feedback and evolving needs.

For the execution of the design approach, we used the Figjam digital whiteboard tool to collect diverse inputs (diverge) and then summarize inputs (converge) in the following steps: Discover, Define, Develop, and Deliver.

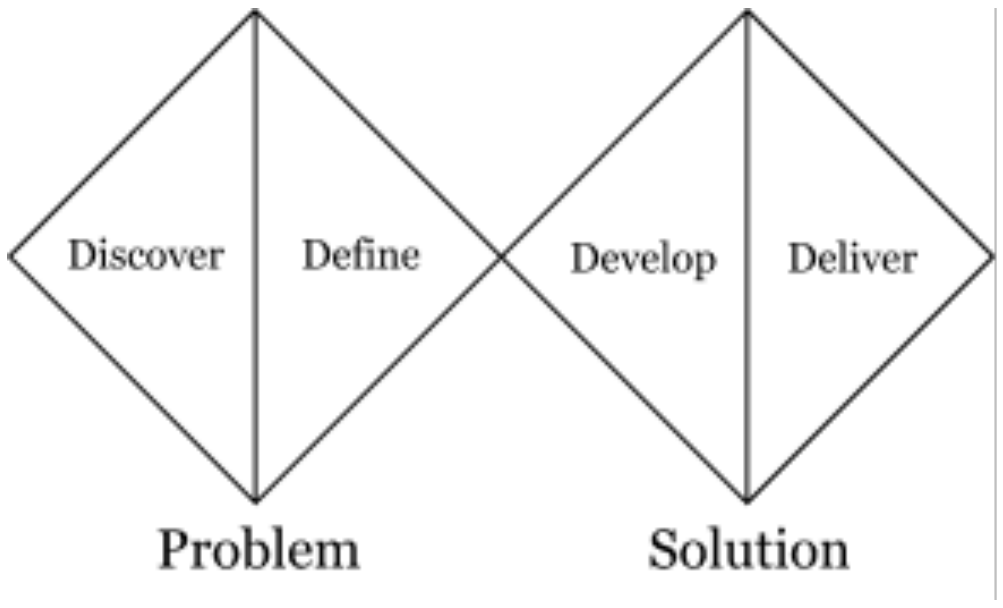

Figure 1 — Double Diamond Problem Solving Framework [15]

The Double Diamond illustrates the process of diverging (considering many possible ideas or solutions) and converging (narrowing down to a manageable set of ideas or solutions), and includes four phases: Discover, Define, Develop, and Deliver [15].

- **Pre-workshop Discovery:** Initial interviews and exploration revealed confusion about the role of the VA. Schemas differed across and within projects, and though dozens of VA fields were available, only a portion of these fields were populated with content. The workshop team also established the main phases of the V&V lifecycle where VAs are used: Planning, Implementation (consisting of Execution, Data Review, and Requirements close out), and Auditing.
- **Discover:** VAs are fundamentally documentation that support planning and collaboration across the V&V program. To "discover" the needs and pain points for that function, for each V&V phase listed above, workshop participants were asked to share the main information needed and decisions that occur.
- **Define:** Participants grouped items to identify themes. They then voted on the most important themes for the MVP schema to support, and the themes that did not receive majority votes were added to a backlog. The workshop team organized the themes into a high-level process, which captured the most important use cases for VAs. At this time, the workshop team also shared prototypes of different approaches that could be supported in Jama so the participants could familiarize themselves with the interactivity and share feedback.
- **Develop:** Ideation consisted of two phases. First, participants matched the most important use cases with candidate VA field and proposed new VA fields as needed. They voted on the most important fields, and the fields that did not receive a majority vote were added to a backlog. Second, the workshop team created a prototype of a VA with the fields agreed on by the majority.
- **Deliver:** Testing consists of two phases. First, the participants walked through the VA prototype field by field, leaving notes. The workshop team either incorporated common concerns into the prototype or added them to the backlog. Second, the schema from the workshops was evaluated as part of pilots in collaboration with the CAE Systems Engineering teams.

In total, 29 individuals participated in four 3-hour sessions spread across 2 months. Their inputs were documented in about 500 unique pieces of information that were categorized, evaluated and dispositioned.

### 3.2 Outcomes

In addition to the VA schema, which is described in detail in Section 4, the workshops produced a implementation-agnostic workflow that captures the main V&V responsibilities across the project lifecycle. Across the V&V phases listed above, practitioners shared a need for Traceability, Results, and Evidence. Scope, Content, and Priorities were particularly important during the Planning phase, as was Schedules, Resources, and Reporting.

### 3.3 Feedback and Reflection

The workshop structure provided an opportunity for diverse V&V practitioners to influence the systems that they use, contribute their wealth of experience and knowledge, and work through disagreements directly. Using voting, grouping, and time-boxing helped the group focus on high-priority, highly relevant topics. Scaffolding the process using the Double Diamond allowed the group to make decisions progressively. In a post-workshop survey, 85% of survey respondents agree that "The final MVP can be used by all of JPL's missions". 100% of respondents agreed that "The MVP emerged from the group's shared understanding of VV at JPL.", and "The workshop format allowed me to share my opinions and feedback". One respondent observed that "There are many, many opinions on how to do VV at JPL - and this effort felt like a really impactful way to get all of us to be advocates for converging on these practices." Another noted that "It may have been better if we started with a baseline MVP and then just did pain points and updates based off of that product rather than starting from scratch." This comment highlights that workshops addressing broad topics with diverse roles can overwhelm the participants with so many inputs. In future workshops it may be more efficient, albeit less transparent, to gather use cases from representative practitioners during workshop preparation and then use workshop time to let participants review the use cases as a group.

The Human-Centered Design Process continues to support the implementation and adoption of the standard schema by designing documentation, conducting follow-up research into how the standard is being used on projects, and supporting updates that emerge from practitioner feedback and the backlog.

## 4. SCHEMA BUILDING BLOCKS

The foundation of the unified V&V Activity standard rests upon a relationship-based paradigm that enhances V&V documentation by treating the requirements management platform as a comprehensive information repository that actively supports planning and collaboration across the V&V program. This approach was established for requirements management at JPL [16] and was extended to encompass the full lifecycle of verification and validation activities.

### 4.1 Integration with Existing Infrastructure

Rather than replacing existing requirements management structures, the V&V schema extends and enhances them. This allows V&V Activities to be associated directly with the defined requirements, enabling multi-directional traceability and analysis.

The schema also accommodates item type selection (allowing projects to choose which method types - such as Test, Analysis, or Inspection - to implement) so that projects may choose specific types based on their needs. This flexibility ensures that the standard can be adapted to varying project contexts while maintaining its core structure. Additional details, such as the templates and Project-Specific Pick Lists further add to the flexibility.

### 4.2 Expansion & Additional Schema Aspects

*Verification/Validation Activities* were included in the previous schema [16], but the approach with one VA was difficult to reconcile with the specific needs of different types and approaches for V&V, as shown in 2.1, for example. This difficulty was also confirmed by the results of the HCDP, which showed that the included information, the specific traceability connections, and even the structures/hierarchies are different between tests and analysis, for example. Thus, after evaluating and scoring various options, a V&V Activity setup that separates the different types was developed.

The schema implements separate item types for different methods, each sharing a common core set of attributes while allowing for method-specific information to be included. This core set of shared attributes formed the foundation for the information architecture described in Chapter 5.

*Venue* represents a critical addition to the schema, specifically designed to support test and demonstration activities. This item type captures the physical or virtual environment where verification/validation is performed, including test facilities, laboratories, simulation environments, and field test locations. By explicitly defining venues as separate items, the schema enables resource planning, schedule deconfliction, and consistent documentation of environmental conditions across multiple test campaigns. The venue concept addresses a key pain point identified during the workshops: the need to track where tests occur and manage facility utilization across projects.

*Relationships* form the critical connective tissue of the V&V schema, establishing explicit links between requirements, activities, and evidence. While the previous schema [16] established fundamental relationships for requirements traceability, the expanded V&V schema introduces specialized relationship types that address the unique demands of verification and validation workflows. These relationships follow a bidirectional philosophy that captures both upstream dependencies and downstream impacts as well as the flow down of information and the flow up of information necessary for V&V completion.

The schema defines several key relationship types with specific directionality rules:

**Verified By** relationships connect requirements to their associated V&V Activities, establishing the foundational traceability between what must be verified and how it will be accomplished. These relationships support both one-to-many mappings (one requirement verified by multiple activities) and many-to-one mappings (one activity verifying multiple requirements).

**Executed In** relationships link test and demonstration V&V Activities to their Venues, documenting where verification/validation was performed and enabling facility usage tracking. This relationship is essential for test repeatability and for understanding the environmental context of results.

Each relationship type carries explicit directionality designations, leveraging upstream/downstream associations to create a responsive network that can propagates change impacts. This approach ensures that when requirements change, the affected V&V Activities can be flagged, and when V&V results indicate issues, the implicated requirements are readily identifiable. Rollup features can also be used to provide summarized status indicators on upper levels, such as VA progress at the requirement level.

The relationship architecture also enables novel analyses previously difficult to perform, such as identifying requirements verified through similar methodologies, locating evidence relevant to multiple requirements, and assessing the V&V status across system elements. By making these relationships explicit rather than implicit, the schema transforms the requirements management platform into a comprehensive information system capable of supporting complex V&V workflows across disciplines and projects.

In addition to the artifact and relationship building blocks, the V&V schema employes various other functions to enable traceability and content structuring without requiring deviation from a standard.

*Outbound references and* links provide a mechanism to handle connections that cross platform and repository boundaries. While the core relationships between requirements and V&V items are managed within the platform, V&V activities frequently interact with external systems for simulation data, test equipment control, and evidence storage. These outbound references maintain traceability to information residing outside the primary repository.

*Templates* for free and rich text fields serve as a remedy for the tension between schema standardization and project-specific customization needs. Templates provide optional content structures that guide text field formatting in a controlled way without mandating rigid structures. This approach allows certain attributes to be outfitted with default templates while enabling customization as needed. The decision whether to set default templates—potentially requiring authors to remove default content if a different format is required—or make all templates optional depends significantly on specific use cases. As detailed in Chapter 5, different decisions were made for certain fields based on user needs identified during the HCDP workshops. Figure 2 below shows an example template in the description field of an Analysis VA.

DESCRIPTION^:

| Associated Analysis Information | |
|---|---|
| Analysis Approach | Describes the methodology used for the analysis, including mathematical models, simulations, or engineering assessments. Example: Finite Element Analysis (FEA) for structural assessment. |
| Inputs and Assumptions | Lists key inputs such as design parameters, boundary conditions, and assumed constraints. Example: Material properties, load conditions, and environmental factors. |
| Analytical Tools and Models | Specifies software, computational models, or theoretical frameworks. Example: MATLAB, ANSYS, or custom Python scripts. |
| Acceptance Criteria | Defines the conditions under which the analysis confirms compliance. Example: Maximum stress should not exceed 80% of yield strength. |
| Results and Findings | Summarizes the output of the analysis, including numerical results, graphical data, and key insights. Example: The structure meets fatigue life requirements under expected operational loads. |

| Record Information | |
|---|---|
| Uncertainty and Limitations | Identifies potential sources of error, assumptions that could impact accuracy, and any limitations of the method. Example: Material properties are based on manufacturer specs and may vary. |
| References & Supporting Documents | Lists relevant standards, technical reports, or prior analyses. Example: NASA-STD-5001, FEA validation reports. |
| Records | Documentation related to the verification, such as calculation logs, tool output reports, and validation checks. Example: Simulation results, verification matrices.<br>URL |

Figure 2 — Template Example

*Project-Specific Pick Lists* provide controlled vocabulary options for categorical fields while maintaining cross-project standardization of the underlying data structure. Unlike free text fields that can lead to inconsistent terminology, pick lists constrain inputs to a predefined set of values, enhancing data quality and enabling more effective filtering and reporting. The V&V schema incorporates two types of pick lists: core lists with standard options applicable across all projects, and extensible lists that projects can customize to reflect their specific nomenclature or methodological approaches. This also considers the fact that exchanges with partner organizations and institutions sometimes cannot be aligned sufficiently to fully standardize all selections and options.

For example, status might use a standard core list (e.g., "Not Started," "In Work," "Complete," "Waived"), while test facility names would use project-specific extensible lists. This approach balances standardization with flexibility, ensuring that categorical data remains structured while accommodating legitimate project variations. The implementation details, including specific fields using pick lists and their extension mechanisms, are further elaborated in Chapter 6.

These building blocks—item types, relationships, outbound references, templates, and pick lists — form a comprehensive schema that addresses the diverse needs identified through the workshops while maintaining a consistent framework across projects and disciplines. By establishing these fundamental elements, the schema provides a solid foundation upon which the detailed information architecture can be constructed. The next chapter expands upon this foundation, detailing the formal data model that transforms these conceptual building blocks into a structured information architecture independent of any specific technology platform—thus making it tool-agnostic.

## 5. Data Model & Information Architecture

While the current schema has been implemented in Jama as described in Chapter 6, its fundamental design is intentionally platform-agnostic. This technology-neutral approach ensures the conceptual architecture can be transferred to different platforms in the future while maintaining clear interfaces and compatibility with complementary systems and technologies. By separating the logical data model from its physical implementation, the schema establishes a durable information architecture that can evolve alongside changing technological ecosystems.

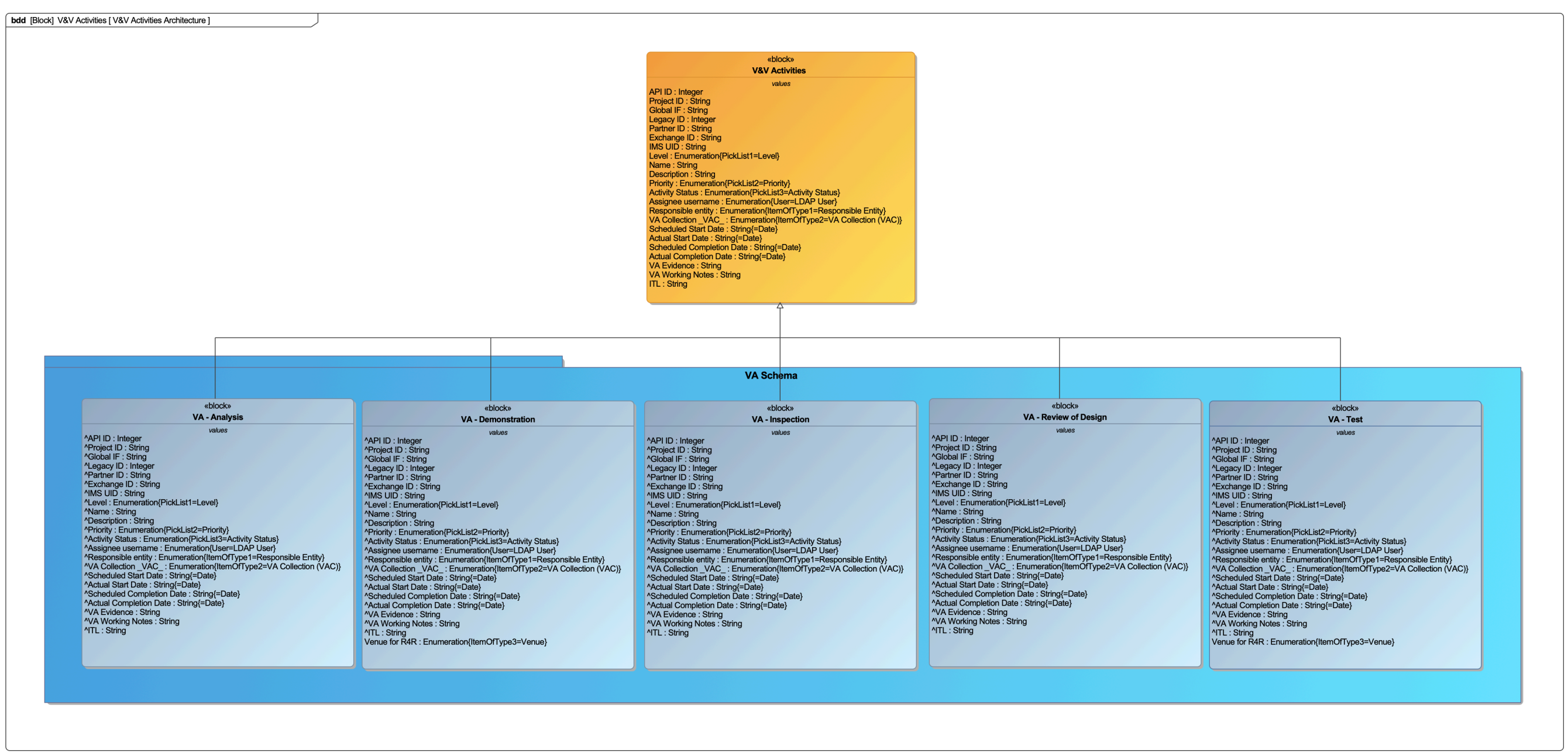

Figure 3 — V&V Activity SysML Model

## 5.1 SysML Information Model

To formalize the V&V Activity schema in a tool-independent manner, a comprehensive SysML model was developed (Figure 3) that captures the essential structure and relationships of activities. This model serves as the authoritative definition of the schema, independent of any particular implementation technology.

As illustrated in Figure 3, the MVP V&V Activities architecture consists of a base MVP (Minimum Viable Product) block that defines common attributes shared across all VA types, with specialized blocks for each method (Analysis, Demonstration, Inspection, Test, and Review of Design).

The inheritance-based structure enables both consistency in core information capture and flexibility in method-specific attributes.

The base MVP block defines universal attributes including activity identifiers, ownership information, status tracking fields, scheduling parameters, and evidence documentation. Each specialized subclass then extends this foundation with method-specific attributes. For example, the Test activity includes venue information, while Analysis activities contain different metadata relevant to analytical approaches.

In addition, the enumerations and picklists visible in the model formalize the controlled vocabularies that ensure consistency across projects while allowing for customization as needed. These enumerated types include Activity Status, Priority levels, and VA Collection types (plus Venue for Run for Record (R4R), where applicable), providing a structured approach to categorical data that maintains semantic consistency while supporting filtering and reporting needs. A more detailed inheritance trace table was also provided for easier understanding of the inheritance and shared attributes—see Figure 4.

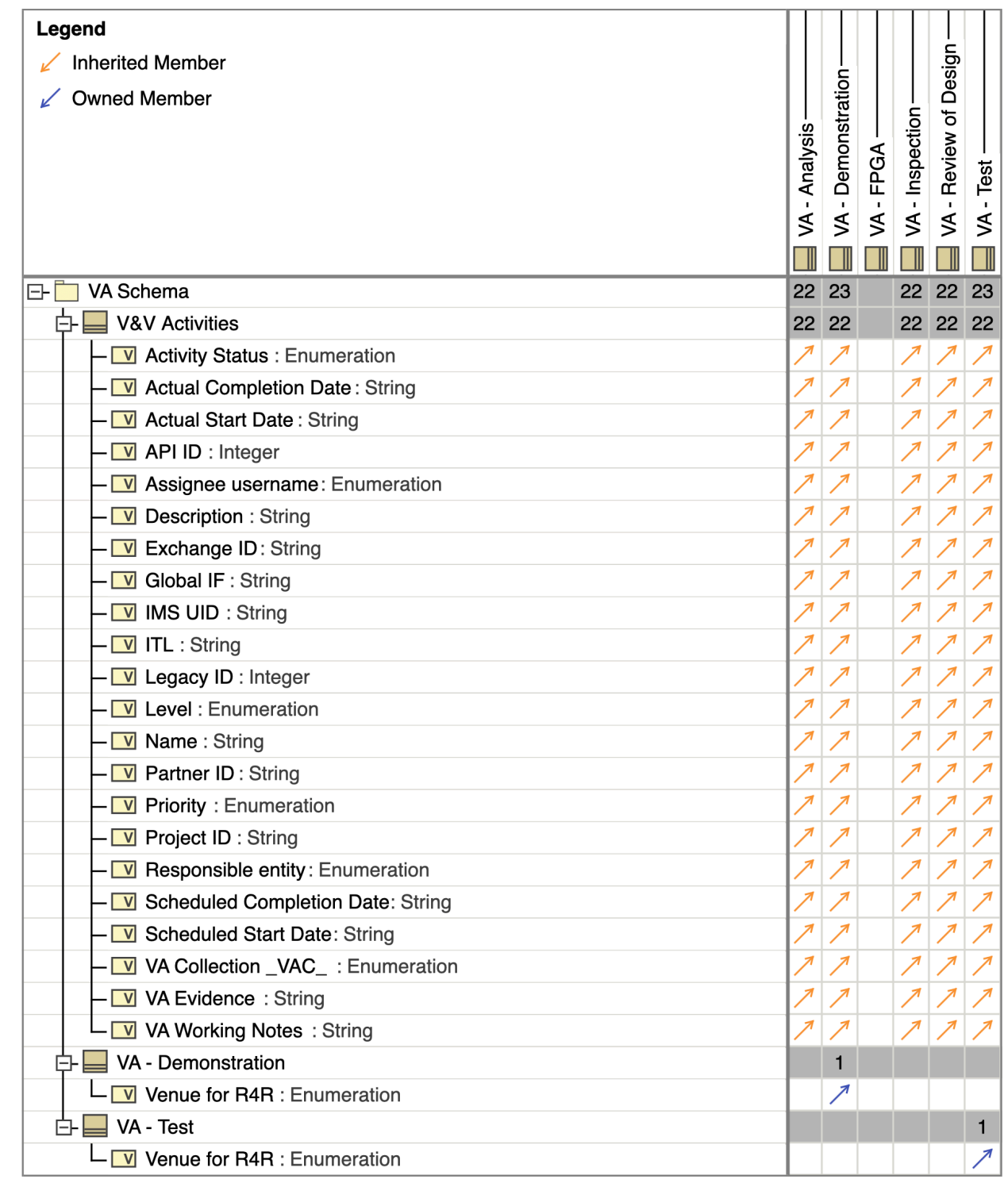
Legend
↙ Inherited Member
↙ Owned Member

| | VA - Analysis | VA - Demonstration | VA - FPGA | VA - Inspection | VA - Review of Design | VA - Test |
|---|---|---|---|---|---|---|
| VA Schema | 22 | 23 | | 22 | 22 | 23 |
| V&V Activities | 22 | 22 | | 22 | 22 | 22 |
| Activity Status : Enumeration | ↗ | ↗ | | ↗ | ↗ | ↗ |
| Actual Completion Date : String | ↗ | ↗ | | ↗ | ↗ | ↗ |
| Actual Start Date : String | ↗ | ↗ | | ↗ | ↗ | ↗ |
| API ID : Integer | ↗ | ↗ | | ↗ | ↗ | ↗ |
| Assignee username : Enumeration | ↗ | ↗ | | ↗ | ↗ | ↗ |
| Description : String | ↗ | ↗ | | ↗ | ↗ | ↗ |
| Exchange ID : String | ↗ | ↗ | | ↗ | ↗ | ↗ |
| Global IF : String | ↗ | ↗ | | ↗ | ↗ | ↗ |
| IMS UID : String | ↗ | ↗ | | ↗ | ↗ | ↗ |
| ITL : String | ↗ | ↗ | | ↗ | ↗ | ↗ |
| Legacy ID : Integer | ↗ | ↗ | | ↗ | ↗ | ↗ |
| Level : Enumeration | ↗ | ↗ | | ↗ | ↗ | ↗ |
| Name : String | ↗ | ↗ | | ↗ | ↗ | ↗ |
| Partner ID : String | ↗ | ↗ | | ↗ | ↗ | ↗ |
| Priority : Enumeration | ↗ | ↗ | | ↗ | ↗ | ↗ |
| Project ID : String | ↗ | ↗ | | ↗ | ↗ | ↗ |
| Responsible entity : Enumeration | ↗ | ↗ | | ↗ | ↗ | ↗ |
| Scheduled Completion Date : String | ↗ | ↗ | | ↗ | ↗ | ↗ |
| Scheduled Start Date : String | ↗ | ↗ | | ↗ | ↗ | ↗ |
| VA Collection _VAC_ : Enumeration | ↗ | ↗ | | ↗ | ↗ | ↗ |
| VA Evidence : String | ↗ | ↗ | | ↗ | ↗ | ↗ |
| VA Working Notes : String | ↗ | ↗ | | ↗ | ↗ | ↗ |
| VA - Demonstration | | 1 | | | | |
| Venue for R4R : Enumeration | | ↗ | | | | |
| VA - Test | | | | | | 1 |
| Venue for R4R : Enumeration | | | | | | ↗ |

Figure 4 — Attribute Trace Table

While the blocks and their relationships define the content and semantic dependencies, a relationship schema was also developed which defines the inter-object relationships as well as information flows. As such, the relationship schema was modeled as a separate Block Definition Diagram. This approach captures not just the static structure of V&V activities but also the dynamic interactions between different elements of the V&V ecosystem. The IBD visualizes how information flows between requirements, V&V activities, evidence products, and other schema elements, providing a comprehensive view of the V&V process that guides implementation decisions across platforms. Figure 5 shows the developed schema, which expands the setup previously defined for the requirement schema [16].

Unlike the information architecture of the items, the relationship schema focuses on the nature of the connections between objects of different types and purposes. Association blocks were used to define not only the linkages and their directionality, but also their semantic naming. As indicated by the association blocks, the direction and nature of each individual relationship prescribe the flow of information within the system. For example, a requirement has a "Verified By" relationship to a VA, meaning that information flows from the VA back towards the requirement, establishing an inverse "Verifies" relationship. Through this bidirectional relationship framework, the information network necessary to close critical items can be constructed and efficiently traversed throughout the development process. This provides complete and seamless traceability not only of individual items, but also of information flow and current state, which is a critical capability for managing complex V&V campaigns across discipline boundaries.

Overall, the model not only documents the schema structure but also serves as a formal specification that can be used to validate implementation conformance across different platforms. The explicit modeling of inheritance relationships and attribute definitions ensures consistency in how V&V information is captured and managed, regardless of the underlying technology stack.

## 5.2 SysML Information Model Incorporation

In addition to the standard way of documentation, the SysML model provides significant advantages by serving as an executable specification that can actively inform implementation across platforms and technologies. By representing the V&V schema in a platform-independent modeling language, we establish a foundation for automated integration and schema evolution.

For one, the model enables API-driven schema updates, allowing future refinements to be propagated systematically across implementations. When modifications are required—whether adding or editing attributes, refining relationships, or introducing new types—changes can be made to the SysML model and then programmatically transformed into implementation-specific configurations. This approach minimizes manual reconfiguration errors and ensures consistency across platforms.

In addition, for interoperability scenarios, the model serves as a contract that defines expected data structures and relationships. External systems can query the schema definition to understand available attributes, valid relationships, and semantic meaning without requiring hardcoded knowledge of the implementation details. This capability is particularly valuable in aerospace environments where V&V data must flow between diverse tools across organizational boundaries. For example, partners and suppliers can use the schema to set up their environments accordingly, which prevents the necessity for mappings and or transformations down the line.

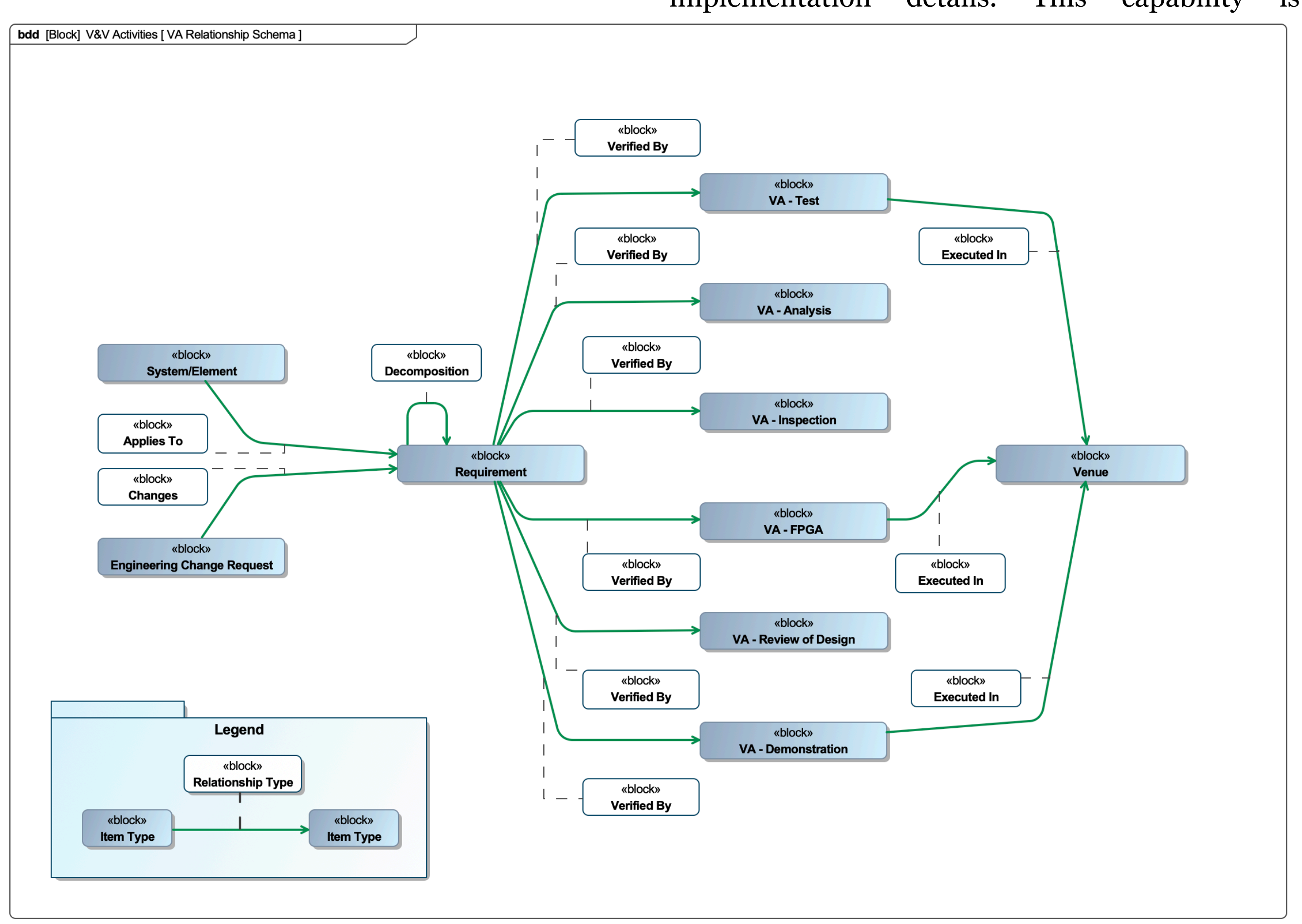

Figure 5 — SysML Relationship Schema Model

In addition, the inheritance-based architecture captured also facilitates extension without modification of the core schema. New activity types can be introduced as subclasses of the base activity, inheriting common attributes while adding specialized properties. This extensibility ensures the schema can accommodate emerging V&V approaches without disrupting existing implementations.

Also, implementation teams can leverage the model for validation purposes, comparing database configurations against the reference model to identify deviations or incomplete implementations. Automated validation scripts can traverse both the model and implementation to generate compliance reports, ensuring that all platforms maintain schema integrity.

Lastly and perhaps most importantly, the model serves as a semantic bridge between technical implementation teams and domain experts focused on V&V processes. By providing a visual representation of the information architecture, the model facilitates meaningful collaboration between stakeholders who may not share common technical vocabularies, but who all participate in the V&V lifecycle even at different stages if needed.

## 6. JAMA IMPLEMENTATION & PRINCIPLES

This section details the practical implementation of the V&V schema within JPL's institutional requirements management platform, Jama. While the information architecture described in Section 5 provides the conceptual foundation, effective implementation requires careful consideration of platform-specific capabilities, limitations, and configuration options to enable the benefits outlined earlier. The implementation of the V&V schema transforms the conceptual information architecture into an operational system supporting activities across JPL, balancing standardization with the flexibility required for diverse mission needs.

### 6.1 Platform & Project Configuration

Building on the standard schema and template developed in accordance with the requirements schema [16], the implementation of the V&V aspects built on top of the standard platform configuration that balances institutional standardization with project-specific flexibility. As such, the core project template—serving as the starting point for all project implementations—was expanded to contain the complete V&V schema. This template includes:

**Base Configuration:** Pre-selected item types, relationship rules, and core pick lists that implement the information model described in Section 5—this is the foundation provided by the standard and further explained in 6.2. The base configuration also includes a standardized hierarchical folder structure in accordance with common systems engineering areas and needs, which was extended to include V&V.

**Permission Setup:** Group-based access controls that align with typical project organizational structures while allowing project-specific refinement. Due to Jama's internal permissions management structures access to specific project areas can be handled within the project while admin and far-reaching permissions are handled on an organization level by the infrastructure providers and discipline representatives.

**Integration Points and Pre-Defined Automations:** Pre-configured connections and scripts working/interfacing with common external systems used in workflows, such as Jira. These automations align with the institutional systems engineering processes and expand standardization capabilities beyond Jama while also further underscoring the benefits and efficiency gains of common information architectures.

Overall, the platform configuration leverages Jama's component-based architecture to create logical separation between global schema/architecture elements maintained at the institutional level and project-specific content and information. How and what the individual projects need to record and document is not defined, but rather framed and guided by the standard.

Project setup follows a standardized process where the base template is instantiated and then tailored to specific project needs through controlled customization, such as specific text field templates or item of type customizations (see project specific picklists below).

## 6.2 Item Type Implementation

The V&V information model was implemented in Jama through a carefully designed set of item types that balance standardization with flexibility:

**Core V&V Activity Types:** Since separate item types were created for each V&V method (Test, Analysis, Inspection, Demonstration, Review of Design), each with method-specific attributes while sharing a common core attribute set derived from the base class in the information model, projects have the option to use or not use the different types they need. For instance, if a project does not have significant partner or supplier connections, Review of Design might not be necessary and can be deselected, in which case it is simply not enable in Jama. This also allows for future changes of additional item types if needed.

**Supporting Item Types:** Additional item types implement supporting concepts such as Venues. Furthermore, project specific picklists are enabled through the use of Jama's "Item of Type Feature" [17]. By referencing existing items within a picklist, a project can define what content they would like to have available for certain attributes, if any. For example, "Responsible Entity" is handled on a name basis by some project and on a role basis by others, which makes it difficult to standardize. By using an "Item of Type" attribute, the definition of the exact picklist items is left up to the project, and the available options are simple managed by creating or deleting "Entity" item types. Moreover, additional information can be kept in the "Entity" Item type, allowing for further customization within the project. Figure 6 shows the functions of "Item of Type" attributes.

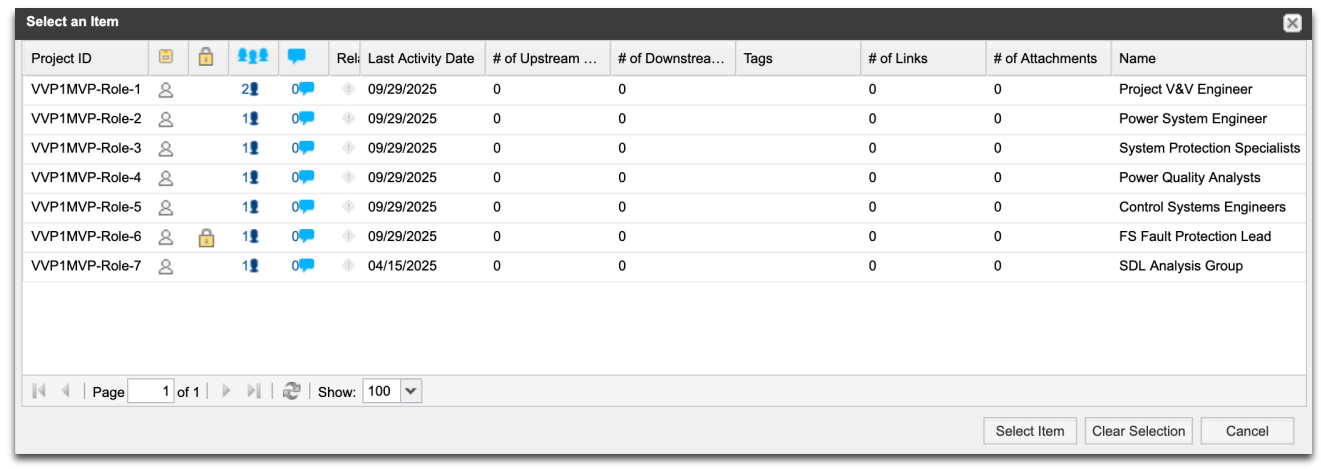

Figure 6 — "Item of Type" Feature and Implementation

**Inheritance Implementation:** While Jama does not natively support inheritance, this concept was implemented through careful attribute replication and naming conventions that maintain the logical class structure. We hope that with future standards, such as SysML v2, even these core concepts can at some point be implemented directly in the technology and application [18].

On a larger scale, the expansion and inclusion of the V&V schema provides a tailorable approach to project setup and content. Unlike requirements, which are typically mandatory across all projects, verification and validation activities have varying applicability depending on mission type, scope, and complexity. The modular architecture leverages Jama's functionality to enable specific item types, processes, and rules on a per-project basis, allowing each project to configure precisely what it needs without overwhelming engineers with irrelevant options.

This approach transforms the standard from a rigid framework into a flexible set of capabilities that projects can selectively implement based on their specific V&V needs and mission characteristics. For example, a CubeSat project might enable a streamlined approach focused on test and inspection, while a flagship mission could implement the full spectrum of V&V methods with extensive evidence management. Figure 7 illustrates this modular implementation concept, showing an example menu that allows different projects to activate different portions of the schema while maintaining consistency in core V&V structures.

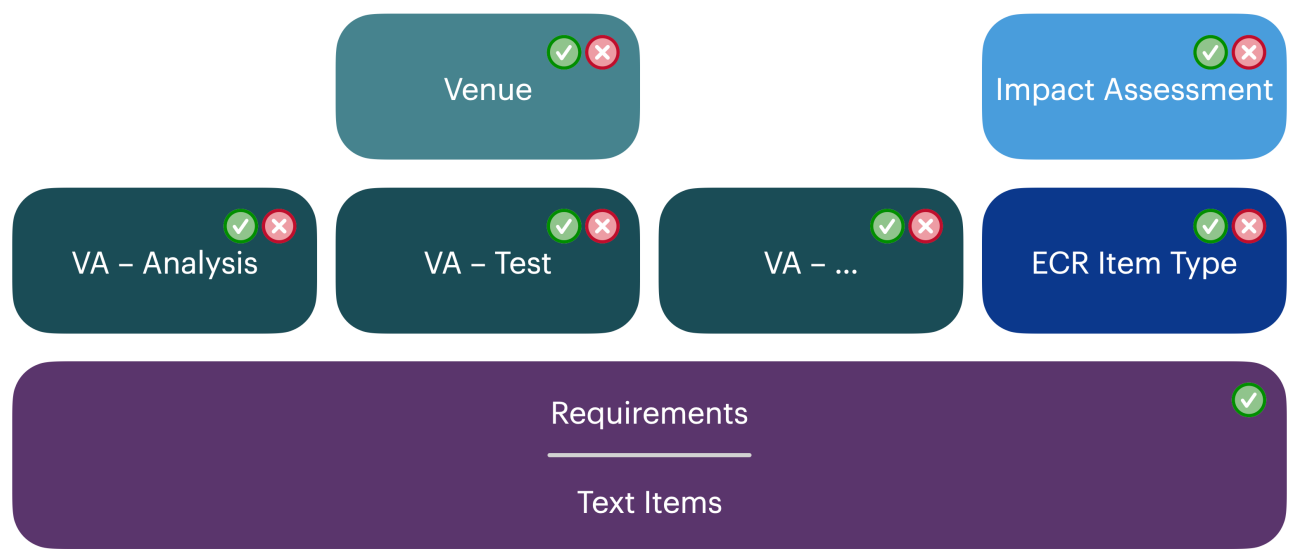

Figure 7 — Scalable and Selectable Framework

## 6.3 Relationships & Traceability

The relationship schema in accordance with the schema shown in Figure 5 implements the bidirectional information flows through Jama's upstream/downstream relationship types:

Core V&V Relationships:

- "Verified By" connects requirements with downstream activities
- "Executed In" links test activities to venues

**Relationship Rules:** Directionality and coverage rules enforce the schema's information flow. Monitoring for missing relationships can highlight V&V gaps through trace views and reports.

**Impact Analysis:** The implementation configures Jama's suspect links feature to propagate change impacts across the relationship network, automatically flagging affected items when related items change.

**Traceability Views:** Custom views enable multi-dimensional navigation through the network, allowing users to trace from requirements through activities, or to explore V&V by venue, method, or system element through filtering.

The relationship implementation creates a complete digital thread through the process, enabling both closure tracking and impact analysis for proposed changes. In addition, the recording of evidence location as part of the information content also enables linking of artifacts in other repositories, in addition to what is described in Section 6.5.

### 6.4 Templates & Content Customization

While maintaining a standardized schema, the implementation accommodates project-specific needs through several controlled extension mechanisms:

**Field Templates:** Rich text fields contain optional templates that provide structure while allowing project-specific content. For example, analysis activities can include templates for assumptions and limitations that guide content creation while permitting customization. Each project can request and design their own templates if they choose to do so and even different templates for the same fields are possible. Due to Jama's full support of HTML, even elaborate tables and formatting options can be included, which can further customize project content. The drawbacks described in 4.2 still apply.

**Extensible Pick Lists:** The implementation distinguishes between institutional pick lists with fixed values and project-specific pick lists that can be extended (see "Item of Type" under 6.2). This approach maintains data structure for core categories while allowing projects to add specialized values.

**View Customization:** Projects can create custom V&V views and dashboards to support specific workflow needs without compromising the underlying data structure.

**Reporting Templates:** Standard report templates can be customized to meet project-specific communication needs while maintaining consistent V&V status reporting.

These customization pathways enable projects to tailor the environment to their specific needs without compromising schema integrity or institutional standardization.

This multi-faceted approach to customization ensures that the schema remains adaptable to diverse project needs while preserving the core standardization benefits. By providing controlled extension points rather than allowing arbitrary schema modifications, the implementation balances flexibility with governance, enabling project-specific workflows without fragmenting the ecosystem. The resulting system maintains institutional consistency while embracing the inherent diversity of approaches across different mission types and engineering disciplines.

### 6.5 Interfaces with External Systems

When it comes to the interface with external systems, two capabilities of Jama played a major role: 1) the feature to directly link outside information in the form of hyperlinks and 2) the API. These capabilities enable Jama to serve as a central hub in a broader ecosystem rather than an isolated repository.

The hyperlink capability allows V&V activities to reference external evidence, documentation, and resources without duplicating data, which would violate the source of truth. This approach maintains traceability while keeping the Jama environment streamlined. For example, test activities can link directly to data repositories containing test results, analysis activities can reference modeling environments, and inspection activities can connect to document management systems containing official records. These connections create a digital thread that spans system boundaries while maintaining Jama as the authoritative source for V&V status and traceability.

Jama's API capabilities enable more sophisticated integrations, allowing bidirectional information flow between systems. The implementation leverages these capabilities to synchronize planning information with schedule management tools, propagate requirement changes to affected activities, and update V&V status based on evidence from external test systems. These integrations reduce manual data entry and ensure consistency across the systems engineering toolchain.

Looking forward, the V&V schema can integrate with future capabilities, particularly through orchestration tools like Syndeia [19]. This systems engineering digital thread platform will enable richer visualization and management of connections between activities in Jama and related artifacts in other repositories. For example, Syndeia will allow teams to track and visualize relationships between Jama activities and corresponding work items in Jira, connecting planning directly to work execution. Similarly, connections to CAD repositories, simulation environments, and PLM systems will provide comprehensive traceability from requirements through design to V&V, creating a complete digital thread across the systems engineering lifecycle. Future integration possibilities can be realized when connections in Syndeia are automatically created and maintained by other cross-platform automations, making the maintenance of a seamless digital thread completely automatic. For example, the creation of test items in an execution environment can initiate a Syndeia link back to the respective VA. Figure 8 below shows an example graph view with Jira and Jama items.

Overall, such integration capabilities—even if just realized through direct hyperlinking—are particularly important for breaking down the discipline silos identified as a key challenge in Section 2, as they enable V&V information to flow seamlessly between the tools used by different engineering specialties while maintaining consistency in approach and status reporting.

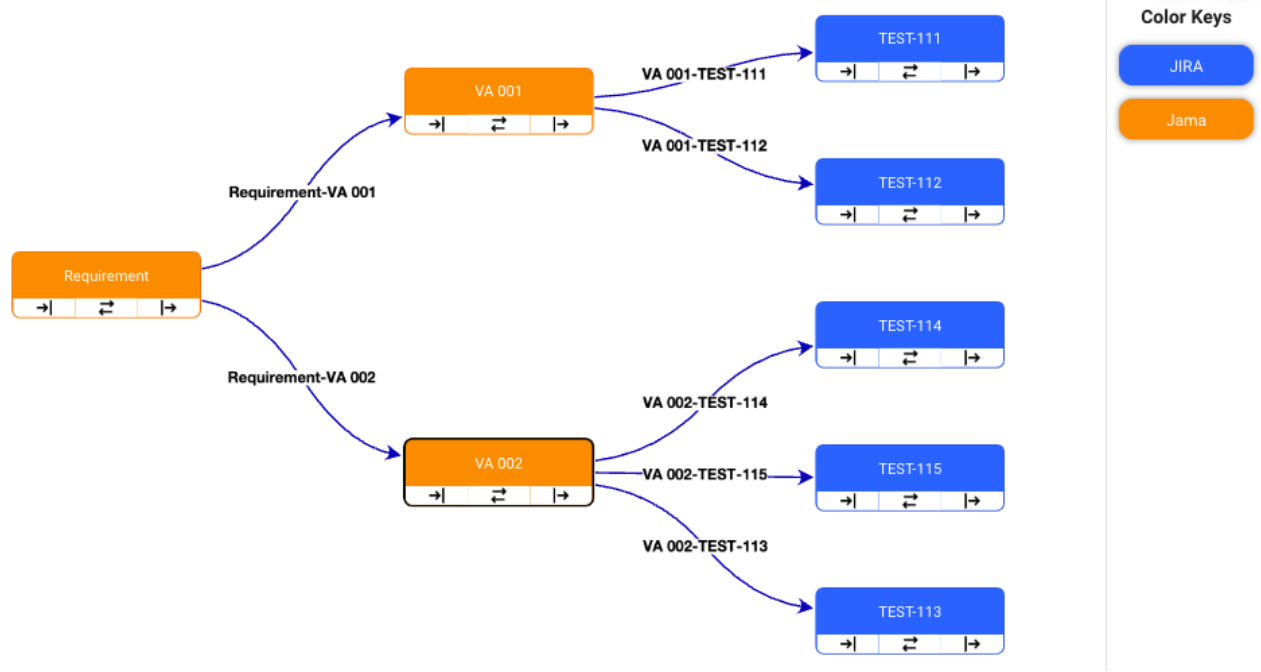

Figure 8 — Linking across Repositories through Syndeia

## 6.6 Current Limitations & Mitigations

Despite the comprehensive implementation approach, several platform limitations required mitigation strategies:

**Concurrence and Synchronization Issues:** A significant limitation is Jama's lack of webhook functionality or outward-facing initiation mechanisms. This creates challenges for maintaining synchronization with external systems in real-time, particularly for bidirectional integrations. To mitigate this limitation, the implementation employs scheduled synchronization jobs and intermediate integration services that poll for and publish changes on a regular basis. While this approach introduces some latency, carefully tuned synchronization intervals minimize the impact on V&V workflows. In addition, on-demand initiated updates are enabled for key project people and administrators.

**Rollup Calculation Constraints:** Jama limits calculation capabilities to data contained within the platform, preventing direct incorporation of metrics or status information from external systems into rollup calculations. This limitation affects comprehensive status reporting that depends on evidence stored in external repositories. The implementation addresses this through a combination of structured outbound references and supplemental reporting tools that aggregate status information across systems. These tools present a unified V&V status view by combining Jama-based metrics with data extracted from linked external systems.

**Document-Based Approval Processes:** Despite the digital transformation objectives, the aerospace industry still relies heavily on document releases and manual signature cycles for formal closure. While exporting V&V information to documents potentially creates source-of-truth ambiguity, it remains necessary for final documentation and sign-off processes. To mitigate this risk, the implementation includes controlled document generation processes with clear traceability back to the authoritative Jama records, along with procedures to ensure consistency between digital and document-based information. Future work will explore digital signature capabilities that may eventually reduce this limitation.

**Inheritance Implementation Challenges:** As noted previously, Jama lacks native support for the inheritance model that forms the foundation of the V&V schema. While attribute standardization and naming conventions maintain the logical structure, this approach introduces maintenance overhead when schema changes are needed. A governance process has been established to manage such changes, ensuring that modifications propagate consistently across all affected item types.

These limitations, while significant, do not fundamentally compromise the benefits of the unified V&V schema. The mitigation strategies ensure that the implementation delivers value despite platform constraints, while ongoing platform evolution may eventually address some of these limitations directly.

The implementation of the V&V schema in Jama demonstrates how a technology-agnostic information model can be effectively translated into an operational system while navigating platform-specific constraints. By carefully balancing standardization with flexibility, the implementation delivers on the promise of breaking down silos across disciplines and projects while accommodating the unique needs of different mission types. Despite certain limitations that required creative mitigation strategies, the resulting system successfully transforms V&V management from disparate approaches into a unified framework that enhances traceability, supports cross-project synergies, and improves efficiency. The next chapter will discuss the migration and transition approach that ensures continuity for existing projects while creating a path toward institutional standardization that will yield increasing benefits as adoption expands across JPL.

# 7. Project Rollout & Implementation

The transition from theoretical schema development to practical implementation across diverse projects represents a critical phase in establishing the unified V&V standard. This chapter explores the initial pilot implementations, lessons learned, and the strategic approach to broader roll-out across JPL projects.

## 7.1 Approach and Goals

The implementation of a new verification and validation schema across an organization like JPL presents significant migration challenges that must be carefully managed. Projects at different lifecycle phases have varying capacities to absorb process changes, with established projects having already committed to specific V&V approaches that are deeply embedded in their planning and execution.

A fundamental challenge lies in the conversion of existing V&V content to the new schema. Such conversions can potentially disrupt traceability and historical information that projects depend on for decision-making if the previous information—or at least references to it—cannot be retained. This risk is particularly acute for projects in critical development or V&V phases where continuity of information is essential for mission success. Further complicating the situation is the reality that multiple legacy schemas exist across JPL due to previous project migrations to Jama, each with their own established V&V structures and processes.

Recognizing these challenges, we established a multi-faceted migration strategy:

**Structured Migration Process:** A formal migration process is being developed in parallel with the standard application procedures. This process includes assessment tools to evaluate migration complexity, phased implementation approaches, and V&V steps to ensure information integrity throughout the transition.

**Lifecycle-Appropriate Adoption:** The migration approach acknowledges that projects at different lifecycle phases require different implementation strategies. Early-phase projects have the flexibility to adopt the new schema completely, while projects in development or operations may implement partial migrations or maintain existing approaches with bridges to the new standard later.

**Preservation of Legacy Schemas (if needed):** To maintain information continuity, existing project schemas can be retained even as the new standard is deployed. This preservation ensures that historical information remains accessible and usable throughout project lifecycles, with carefully managed bridges between legacy and new schema elements where appropriate. In practice, this means that projects have the option to temporarily use more than one schema for V&V while information is being moved from one to the other.

**Early Engagement with Key Projects:** Selected pilot projects were engaged immediately for User Acceptance Testing (UAT), providing valuable feedback for schema refinement while establishing exemplars for future implementations. These projects, chosen to represent diverse mission types and V&V approaches, serve as pathfinders that demonstrate the schema's adaptability and benefits.

**Adoption Incentives:** To encourage voluntary adoption, the standard schema comes with an expanding portfolio of automatic integrations, automation capabilities, and solution accelerators that are designed to work seamlessly with the standard structure. These additional capabilities provide tangible efficiency benefits that motivate projects to adopt the standard even when not explicitly required. This also ties into the future work described in Chapter 8 since it is an ongoing effort.

**New Project Standard:** Lastly, all new projects beginning after the schema deployment will start with the standard V&V schema as their foundation, ensuring consistent adoption moving forward. This approach gradually increases the proportion of projects using the standard while minimizing disruption to existing work.

The above-described migration strategy balances the imperative for standardization with the practical realities of project constraints, creating a path toward unified V&V practices without disrupting ongoing mission development. The approach acknowledges that full standardization is not instantaneous and must align with project lifecycles rather than forcing immediate compliance across the institution. Over time, the different existing schemas will converge and move to the standard, leading to a gradual unification over time as visualized in Figure 9 below.

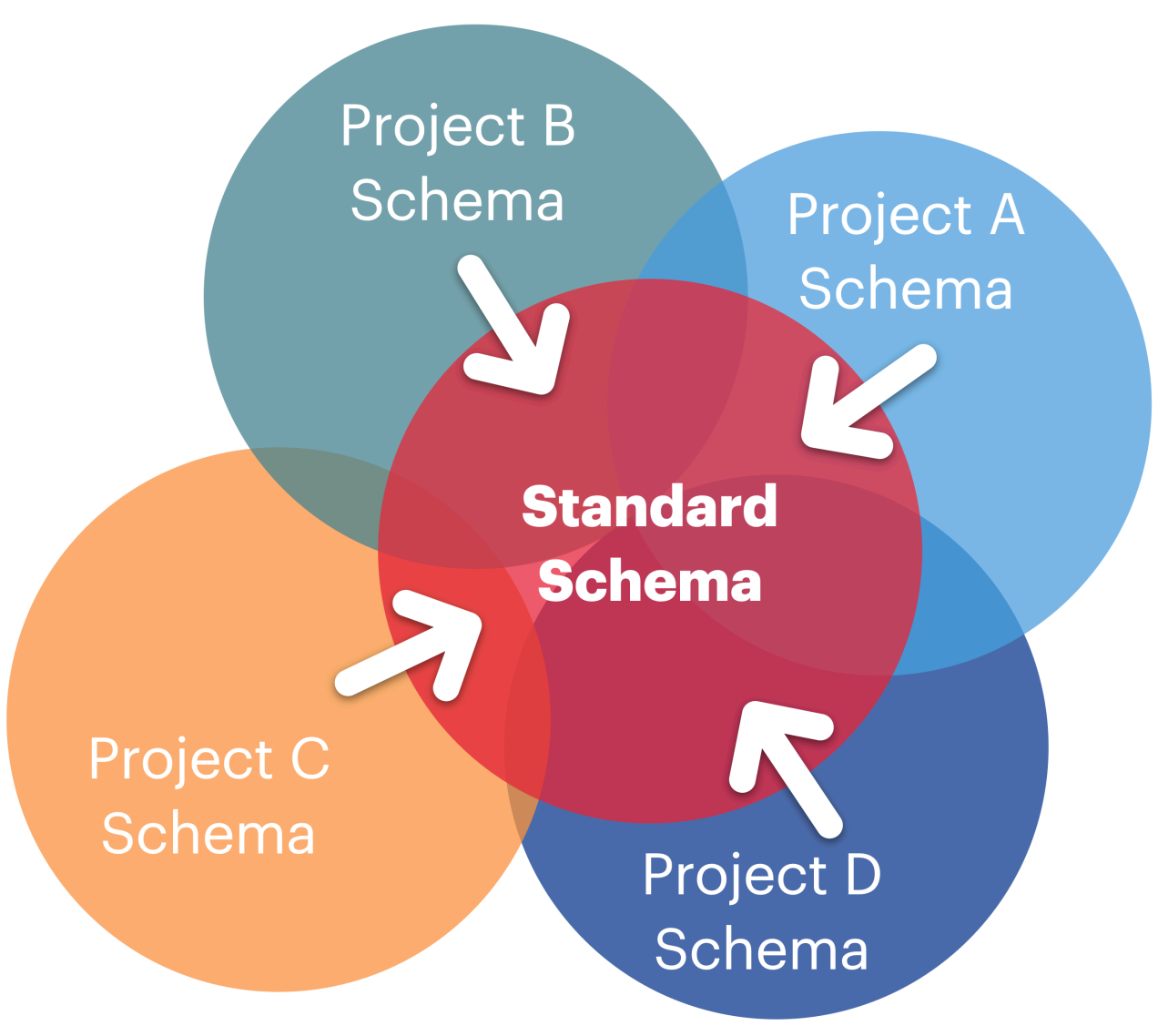

Figure 9 — Planned Progressive Schema Unification

## 7.2 Roll-Out, Status, and Planned Steps

Following the strategic approach outlined in section 7.1, the roll-out of the unified V&V schema has progressed through several key phases. This section outlines the current implementation status and the approach for migrating legacy activities to the new standard.

As mentioned, for new projects, the schema is implemented directly as part of project setup, allowing immediate access to the full capabilities of the unified approach. As such, the schema is also included in the default project template.

For existing projects with established V&V activities, a methodical migration process preserves data integrity by first capturing a complete baseline of existing VAs, then establishing the new item type structure in the project's Jama environment. Using a controlled export-import process, activities are extracted, transformed to map legacy attributes to the new schema, and imported into the new structure while maintaining all relationships—effectively enabling a conversion.

Mapping requires careful consideration of each attribute's meaning and structure. Certain fields must account for different workflow states and others must reconcile different scales. Furthermore, for some, assignments previously handled in free text fields must transition to structured relationships and "Item of Type" fields.

Current implementation status shows promising adoption. New projects have launched with the standard schema, while pilot migrations have validated the approach for existing projects. These cases confirm effectiveness in addressing cross-discipline coordination challenges.

For projects considering migration, an assessment list evaluates complexity based on factors like volume, custom attribute count, and integration requirements. This assessment ensures realistic planning and appropriate support. Figure 10 below shows the high-level steps of the schema migration.

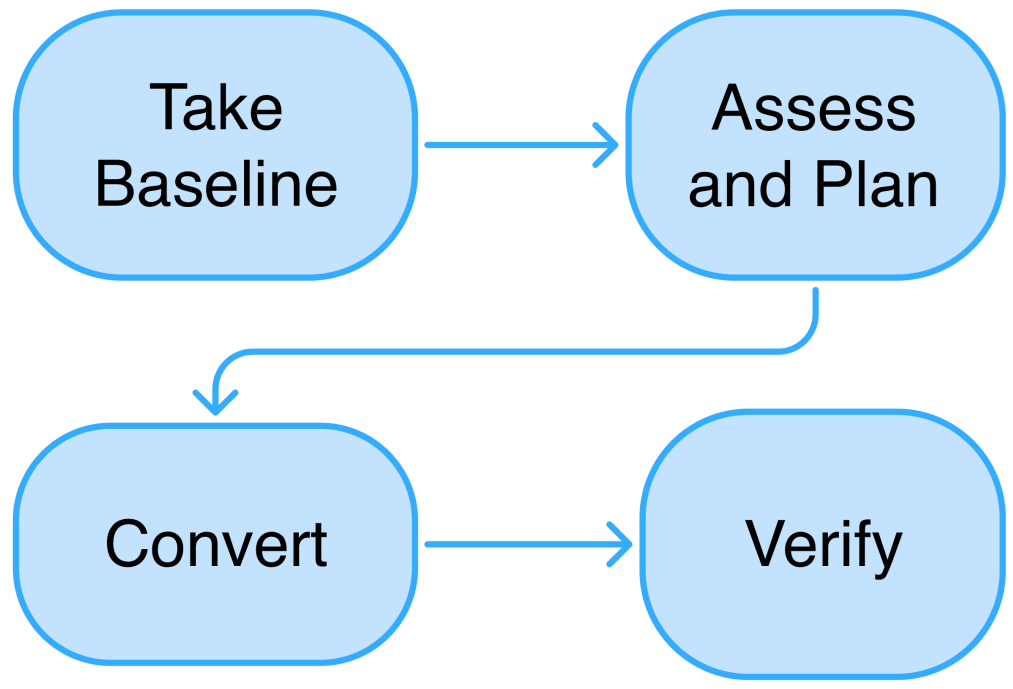

Figure 10 — Schema Migration Process Overview

The roll-out continues with a phased approach aligned with project lifecycles. Near-term plans include expanding adoption, refining migration steps, and developing additional integrations. As more projects adopt the standard, the benefits of cross-project learning and pattern reuse will further accelerate adoption.

Implementation experience—also considering platform migrations such as DNG to Jama—has validated the schema's core design principles while revealing opportunities for refinement. The careful balance between standardization and flexibility has proven essential, allowing projects to tailor their approach within a structured and integrated framework.

## 8. Summary and Future Work

### 8.1 Summary

The development and implementation of a unified Verification and Validation (V&V) Activity standard at JPL represents a significant advancement in systems engineering practice for mission-critical aerospace projects. This paper has detailed the creation of a standardized yet flexible schema that breaks down disciplinary silos while maintaining the rigor essential for mission success. Through human-centered design involving practitioners across disciplines, we have established a framework that balances standardization with the adaptability required for diverse mission contexts.

Key contributions of this work include the development of a relationship-based V&V schema that connects requirements, activities, and evidence through explicit relationships. By modeling the schema in SysML, we created a platform-independent specification that can evolve and be implemented across different technologies. The implementation in Jama demonstrates how this conceptual model translates into operational practice, with carefully designed item types (including respective attributes), relationship rules, and controlled customization pathways.

The standard addresses critical challenges identified during practitioner workshops, including cross-discipline coordination, resource optimization, and evidence management. By separating V&V activities by method while maintaining a common core structure, the schema accommodates the distinct needs of different approaches while enabling consistent reporting and analysis. The addition of venue tracking, project-specific pick lists, and templates further enhances the standard's utility across diverse project contexts.

Initial deployments and user tests have validated the approach, confirming that standardization does not need to come with a loss in flexibility. The phased implementation strategy ensures that projects can adopt the standard without disruption, gradually building a consistent V&V ecosystem across JPL while respecting the constraints of ongoing missions.

Through this work, we have demonstrated that rigorous V&V practices can coexist with the agility demanded by modern aerospace development. The unified standard lays the groundwork for continued improvements in V&V efficiency, cross-project learning, and ultimate mission success.

### 8.2 Benefits & Insights

The implementation of the unified V&V Activity standard yields several significant benefits across JPL projects that adopt it:

First, the standardized schema enhances cross-discipline collaboration by providing a common language and framework for VAs . Engineers from different disciplines can now more effectively coordinate their V&V efforts, with clear visibility into dependencies and shared resources. This is particularly valuable for complex V&V scenarios that span multiple engineering domains.

Second, the relationship-based architecture improves requirements traceability and impact assessment. When requirements change, affected VAs can automatically be flagged, enabling more proactive planning. This bidirectional traceability reduces gaps and overlaps, ensuring more comprehensive coverage while minimizing redundant activities.

Third, the separation of methods into distinct but related item types has allowed for more targeted and efficient planning. Teams can more easily identify similar approaches across different system elements, facilitating resource optimization and test consolidation. The venue tracking capability further enhances this benefit by enabling more efficient facility utilization and planning thereof.

Fourth, significant time savings in documentation and report generation are expected as the standardized templates and structured information architecture reduce the effort required to generate documents, while the consistent structure has improved readability and comprehensibility for reviewers and stakeholders.

Fifth, the standardized schema has enabled a foundation for developing universally applicable automations and integrations that benefit the entire JPL community. Because all projects implementing the standard share the same information architecture, automation solutions developed for one project can be immediately deployed to others without modification. This is part of a larger inner-source approach at JPL where projects contribute automation solutions back to the community, creating a virtuous cycle of continuous improvement. As more projects adopt the standard, this shared repository of automation capabilities continues to grow, reducing duplicate effort and accelerating activities across the institution. Such automations and integrations can especially help smaller projects that may not have the capacity or the funds to develop solutions themselves.

Finally, early adopters have noted the easier facilitation of knowledge transfer between projects since a common framework makes it easier to understand and adapt V&V approaches from previous missions, accelerating V&V planning and reducing rework. As adoption continues to expand, this cross-project learning benefit is expected to grow substantially.

These benefits demonstrate that the standardization approach can enhance effectiveness and efficiency while maintaining the flexibility needed for diverse mission contexts.

### 8.3 Future Work & Pending Efforts

The development of the unified V&V Activity standard establishes a foundation for several promising research and implementation directions. Building on the current schema and implementation, we identify several key areas for future exploration and enhancement.

First, as adoption expands across JPL, opportunities emerge for developing standardized patterns that can be reused across projects. These patterns would capture common approaches for recurring engineering challenges, enabling new projects to leverage institutional knowledge rather than recreating V&V strategies from scratch. By analyzing VAs across multiple missions, we can identify high-value patterns for standardization, develop template implementations, and create a library of proven V&V approaches.

Second, the relationship-based architecture of the V&V schema creates opportunities for advanced analytics and visualization capabilities. By mining the growing database of relationships, we can develop insights about coverage patterns, resource utilization trends, and common V&V challenges. Visual analytics tools that leverage graph theory could provide systems engineers with powerful new ways to understand complexity and optimize V&V planning. In addition, such analytics can be expanded across projects since information and content of different projects can now be compared and analyzed collectively.

Third, a significant advancement on our roadmap is the direct mapping of SysML architecture models to implementation environments through integration platforms like Syndeia. This capability will eventually enable fully automated deployment of schema changes and updates directly from the authoritative SysML models to Jama and other platforms without manual configuration or implementation steps. Once all involved tools mature their API capabilities and model transformation features, we anticipate being able to maintain a single source of truth in the SysML models with automated synchronization to implementation environments. This would dramatically reduce schema maintenance overhead, eliminate implementation discrepancies, and ensure that theoretical models and practical implementations remain perfectly aligned and versioned.

Fourth, while the current implementation includes basic interfaces with external systems, there are significant opportunities to develop deeper integrations with the broader systems engineering ecosystem. Future work will explore real-time synchronization between the V&V management platform and test execution systems, analytical tools, and configuration management databases. These integrations further reduce manual data entry, improve consistency, and create a more seamless digital thread through the entire lifecycle.

Fifth, the platform-independent information model provides a foundation for exploring model-based approaches to planning and execution. By connecting the schema with system architecture models, we could develop capabilities for automatic generation of activities based on design changes, model-based assessment of V&V coverage, and simulation-based V&V planning. These capabilities would further accelerate the lifecycle while maintaining rigor.

Finally, as more projects adopt the standard schema, we see opportunities to develop machine learning and Artificial Intelligence (AI) approaches that leverage the growing repository of V&V data. These approaches could assist in estimating effort, predicting issues based on requirement characteristics, and recommending V&V strategies based on historical patterns as well as project data. By transforming historical data into actionable guidance, these capabilities would further enhance efficiency and effectiveness at large scales.

Together, these future directions represent a trajectory toward increasingly intelligent and integrated V&V management that maintains JPL's commitment to mission success while embracing the innovations needed for next-generation aerospace development. The unified schema presented in this paper forms not just another milestone, but a critical cornerstone in building the road toward a fully integrated systems engineering environment—one where information flows seamlessly across disciplines, tools speak a common language, and the full power of collective expertise can be brought to bear on humanity's most ambitious exploratory missions.

## ACKNOWLEDGMENT

The authors would like to thank all participants and contributors to the workshop as well as all involved stakeholders and function owners. The research was carried out at the Jet Propulsion Laboratory, California Institute of Technology, under a contract with the National Aeronautics and Space Administration (80NM0018D0004).

## REFERENCES

[1] K. Forsberg and H. Mooz, "The Relationship of System Engineering to the Project Cycle," *Center for Systems Management*, vol. 5333, pp. 4-6, 1991.

[2] K. Forsberg, H. Mooz, and H. Cotterman, *Visualizing Project Management*. John Wiley & Sons, 2005.

[3] *Systems and software engineering – System life cycle processes*, 15288:2023, ISO/IEC/IEEE, 2023.

[4] *NASA Systems Engineering Handbook*. NASA, 2019.

[5] A. Salado and H. Kannan, "A mathematical model of verification strategies," *Systems Engineering*, vol. 21, no. 6, pp. 593-608, 2018, doi: https://doi.org/10.1002/sys.21463.

[6] INCOSE, *Systems Engineering Handbook*, 5th ed. Wiley, 2023.

[7] A. Engel, *Verification, Validation, and Testing of Engineered Systems*. John Wiley & Sons, 2010.

[8] D. M. Buede and W. D. Miller, *The Engineering Design of Systems: Models and Methods*, 3rd ed. John Wiley & Sons, 2016.

[9] W. W. Royce, "Managing the Development of Large Software Systems," in *9th International Conference on Software Engineering*, 1970: IEEE.

[10] O. Gotel and A. Finkelstein, "Modelling the Contribution Structure Underlying Requirements," presented at the 1st International Workshop on Requirements Engineering: Foundations for Software Quality, 1994. [Online]. Available: https://openaccess.city.ac.uk/id/eprint/26460/.

[11] O. Gotel and A. Finkelstein, "An Analysis of the Requirements Traceability Problem," in *Proceedings of IEEE International Conference on Requirements Engineering*, 18-22 April 1994 1994, pp. 94-101, doi: 10.1109/ICRE.1994.292398.

[12] *Institutional Project Verification and Validation Plan, Rev. 1*, Jet Propulsion Laboratory and California Institute of Technology, 2021.

[13] D. Pauly, "Anecdotes and the shifting baseline syndrome of fisheries," *Trends in Ecology and Evolution*, vol. 10, pp. 430-430, January 01, 1995 1995, doi: 10.1016/s0169-5347(00)89171-5.

[14] *Ergonomics of human-system interaction*, 9241-110:2019, ISO, 2019.

[15] D. Raposo, J. Neves, and J. Silva, *Perspectives on Design II: Research, Education and Practice*. Springer International Publishing, 2021.

[16] B. Oaida, E. Bovre, N. Blackway, and A. Eikanas, "A Relationship-Based Schema for Requirements Architecting and Management using Jama at JPL," in *2024 IEEE Aerospace Conference*, 2-9 March 2024 2024, pp. 1-15, doi: 10.1109/AERO58975.2024.10520996.

[17] Jama Software. "Custom fields." https://help.jamasoftware.com/ah/en/administration/organization-administrator/managing-process/fields/custom-fields.html (accessed September 24, 2025).

[18] OMG. "SysML version 2 (v2) " https://www.omg.org/sysml/sysmlv2/ (accessed September 24, 2025).

[19] Intercax LLC. "Syndeia® - The Digital Thread Platform for Digital Engineering." https://intercax.com/products/syndeia (accessed September 25, 2025).

## BIOGRAPHIES

**Maximilian Vierlboeck** is a Systems Engineer and Product Owner for Enterprise AI & Requirements Platforms at NASA's Jet Propulsion Laboratory (JPL). He leads the design, implementation, and integration of advanced digital engineering frameworks that enhance requirements and systems engineering across missions and projects. His work encompasses enterprise-scale infrastructure and governance, serving hundreds of engineers and supporting digital engineering initiatives.
Maximilian holds a Ph.D. from Stevens Institute of Technology and previously earned bachelor's and master's degrees in mechanical engineering from the Technical University of Munich (TUM). Before his time at JPL, he gained practical experience in product development with Knorr-Bremse. Maximilian's research has included topics such as complexity assessment frameworks, natural language processing for requirements engineering, model-based systems engineering (MBSE), and agile methodologies. He has been recognized with multiple awards, including for his contributions to the Psyche mission, as well as for research excellence, publications, and academic achievements.

**Ellen Van Wyk** is a Product Designer with over 10 years of experience modernizing workflows and enabling more science at NASA JPL, for academia, and in industry. She has led user experience design for projects across the flight project lifecycle, including Europa Clipper operations, operations for autonomous spacecraft, and digital engineering systems. She is most excited about visual storytelling and creating user-centered experiences for technical users and teams at the cutting edge of technology. She has a master's degree in Information Systems from UC Berkeley and a B.S. in Neurobiology from the University of Washington.

**Natalia Sanchez** is a Flight Systems Engineer and Group Supervisor at NASA's Jet Propulsion Laboratory (JPL) with over 15 years of experience in the verification, validation, and cross-cutting systems engineering of Planetary Science and Earth-observing missions. A graduate of Cal Poly San Luis Obispo, she currently leads strategic initiatives for improvement and innovation of the Systems Engineering practice, including GenAI adoption, within large-scale engineering organizations. Beyond her work at JPL, she is actively engaged in Yogic science and cultivates her interest in philosophy and organizational leadership as complementary disciplines to Engineering.

**Bogdan Oaida** is a Systems Engineer and Technical Group Supervisor at the Jet Propulsion Laboratory (JPL), where he leads the Instrument Systems Engineering Group supporting more than a dozen NASA flight instruments from development through operations. Over his 15 years at JPL, he has served in key systems engineering roles across technology demonstrations, Earth science, and planetary missions—including OPALS, Europa Clipper, and EMIT—spanning concept development, requirements and V&V architecture, mission design, integration and test, and on-orbit operations. He has pioneered adoption of modern requirements and systems engineering tools at JPL, developed standardized processes and lifecycle products, and led major cross-center collaborations with NASA centers and industry partners. In the workplace, he is passionate about data visualization and about bringing together people, processes, and tools to improve how teams design and operate spaceflight instruments.

**Torrance Eberhart** is a Systems Engineer at NASA's Jet Propulsion Laboratory (JPL), where he implements digital engineering frameworks to enhance requirements and systems engineering across multiple projects. He develops Requirements Maturity dashboards that provide critical metrics for mission planning and execution. As part of the Deep Space Optical Communications (DSOC) project, he produced telemetry visualizations and processed data products for the flight operations team. Torrance is pursuing his master's degree in Space Systems Engineering at Johns Hopkins University and holds a bachelor's and a master's degree in mechanical engineering from North Carolina A&T State University. He plans to pursue a Ph.D. focusing on exoplanet instrumentation. He was recognized by NASA for his contributions to DSOC Flight Operations.

**Marie Piette Gomez** is a Software Systems Engineer at the Jet Propulsion Laboratory (JPL). She is the Systems Environment Lead of the Computer-Aided Engineering project, where she leads the team to delivering an integrated systems infrastructure for engineers across the Engineering and Science Directorate. Her primary role is to adapt this environment to the specific needs of various Flight Projects, most recently Mars Sample Return/Sample Retrieval Lander and Mars2020.

**Christopher (Chris) Delp** is a prominent leader at NASA's Jet Propulsion Laboratory (JPL), and a Caltech instructor. As a foremost authority on Digital Engineering and Model-Based Systems Engineering (MBSE), he has been instrumental in transitioning JPL from traditional document-centric workflows to advanced digital engineering practices. He is a co-founder of OpenMBEE, an open-source platform and community that serves digital engineering teams and collaborative engineering models. His extensive portfolio includes leading the digital environment for the Europa Clipper mission and contributing to the global development of SysML standards through the Object Management Group. With over 25 years of experience, Delp remains a central figure in re-architecting how safety-critical flight projects are designed and managed at the Lab.

**Marijke Jorritsma** is a Human-Centered Design Lead and Researcher at the NASA Jet Propulsion Laboratory (JPL), where she specializes in the intersection of Human-Centered Design (HCD) and complex aerospace systems. Her work primarily focuses on developing intuitive interfaces and robust methodological frameworks for mission-critical operations.